\documentclass[12pt]{article}
\usepackage{amssymb,amsmath,epsfig}

\begin{document}
\title{\bf Stability of Anisotropic Cylinder with Zero Expansion}

\author{M. Sharif$^1$ \thanks{msharif.math@pu.edu.pk} and M. Azam$^{1,2}$
\thanks{azammath@gmail.com}\\
$^{1}$ Department of Mathematics, University of the Punjab,\\
Quaid-e-Azam Campus, Lahore-54590, Pakistan.\\
$^{2}$ Division of Science and Technology, University of Education,\\
Township Campus, Lahore-54590, Pakistan.}

\date{}

\maketitle
\begin{abstract}
We study the dynamical instability of anisotropic collapsing
cylinder with the expansion-free condition, which generates vacuum
cavity within fluid distribution. The perturbation scheme is applied
to distinguish Newtonian, post-Newtonian and post-post Newtonian
terms, which are used for constructing dynamical equation at
Newtonian and post-Newtonian regimes. We analyze the role of
pressure anisotropy and energy density inhomogeneity on the
stability of collapsing cylinder. It turns out that stability of the
cylinder depends upon these physical properties of the fluid, not on
the stiffness of the fluid.
\end{abstract}

\section{Introduction}

The study of relativistic anisotropic stars is important due its to
various applications in astrophysical scenarios. The existence of
anisotropy in the star models is justified from physical phenomena
like solid core (Kippenhahn and Weigert 1990), phase transition
(Sokolov 1980), mixture of two fluids (Letelier 1980), slow rotation
(Herrera and Santos 1995) and pion condensation (Sawyer 1972). Some
authors (Lobo 2006; Sharma and Maharaj 2007; Thirukkanesh and
Maharaj 2008; Sharif and Bhatti 2012) have also investigated the
physical significance of charged self-gravitating configurations
with anisotropic pressures.

It is believed that the expansion scalar characterizes the motion of
fluid. Since it describes small change in the volume of the matter
with respect to proper time, so it has a significant role in the
evolution of the relativistic systems. In this scenario, Skripkin
(1960) presented an expansion-free model for the non-dissipative
isotropic fluid. Herrera et al. (2008) generalized this to
anisotropic and dissipative fluid by defining the motion of
expansion-free fluid with radial velocity in two different ways. The
study of such models, where the expansion-free condition necessarily
generates a vacuum cavity within the fluid distribution has been of
particular interest. The evolution of cavities has also been shown
under purely areal evolution condition (Herrera et al. 2010). The
same authors (Herrera et al. 2009) found that Skripkin model for
non-dissipative perfect fluid with constant energy density is
inconsistent with junction conditions and concluded that in general
for such a fluid distribution the expansion-free condition requires
inhomogeneous energy density. Di Prisco et al. (2011) studied the
spherically symmetric distribution of anisotropic fluid with
inhomogeneous energy density and found some exact analytic models
under expansion-free condition.

In general relativity (GR), the stability of self-gravitating
objects against perturbation is an important issue. The existence of
any static configuration has been constrained through its stability
against perturbations. Some people (Ivanov 2002; Dev and Gleiser
2003; Mak and Harko 2003) have investigated the role of anisotropy
for the stability of physical objects. Also, different physical
properties of fluid such as heat conduction, radiation density and
viscosity would increase or decrease the instability range of the
system (Herrera et al. 1989; Chan et al. 1989; 1993; 1994). Horvat
et al. (2011) explored the stability of anisotropic configurations
under radial perturbations. Sharif and Kausar (2011) have studied
dynamical instability of the expansion-free fluid in $f(R)$ gravity.
Herrera et al. (2012a) found that instability of the expansion-free
fluid depends exclusively on energy density inhomogeneity and
pressure anisotropy. In recent papers, we have also explored the
problem of dynamical instability of cylindrical (Sharif and Azam
2012a), spherical (Sharif and Azam 2012b) and thin-shell wormholes
(Sharif and Azam 2012c).

Recent contributions (Di Prisco et al. 2007; Sharif and Bashir 2012;
Kalam et al. 2012; Hossein et al. 2012) indicate substantial
importance of anisotropic pressure on the dynamical evolution of
relativistic objects. The role of cylindrical gravitational waves in
GR allows to study the dynamics of cylindrical geometry. For
instance, investigation of naked singularity (Nolan 2002), emission
of gravitational radiation (Sharif and Ahmad 2007) from cylindrical
gravitational collapse, gravitational collapse of hollow cylinder
(Nakao et al. 2009) and structure scalars for cylindrically
symmetric metric with dissipative anisotropic fluid (Herrera et al.
2012b). Recently, Sharif and Yousaf (2012) have found some exact
analytical expansion-free cylindrical symmetric models with
anisotropic fluid. This shows that anisotropy is important in
astrophysical process.

In this paper, we investigate the dynamical instability of
cylindrical symmetric spacetime with expansion-free condition. The
paper has the following format. Section \textbf{2} deals with
properties of kinematical variables of the fluid, Einstein field
equations and junction conditions. In section \textbf{3}, we perturb
the field equations, dynamical equations and the mass function.
Section \textbf{4} describe the dynamical instability of
expansion-free fluid at Newtonian and post Newtonian (pN) regimes.
We summarize the results in the last section.

\section{Matter Distribution of Collapsing Cylinder}

In this section, we describe some basic kinematical variables
associated with the collapsing cylinder in the comoving coordinate
system. For this purpose, we consider a cylindrical geometry in
the interior region given as (Sharif and Azam 2012a)
\begin{equation}\label{1}
ds^2_-=-A^2(t,r)dt^{2}+B^2(t,r)dr^{2}+C^2(t,r){d\theta^{2}}+ dz^2,
\end{equation}
where $-\infty{<}t{<}\infty,~0\leq{r}<\infty,~
-\infty<{z}<{\infty},~0\leq{\theta}\leq{2\pi}.$ Here we assume the
fluid to be locally anisotropic. The energy-momentum tensor for such
a fluid has the form (Sharif and Yousaf 2012)
\begin{equation}\label{2}
T^-_{\alpha\beta}=(\mu+p_r)v_{\alpha}v_{\beta}+p_{r}g_{\alpha\beta}
+(p_z-p_r)s_{\alpha}s_{\beta}+(p_\theta-p_{r})k_{\alpha}k_{\beta},
\end{equation}
where $\mu,~p_{r},~p_{\theta},~p_{z}$, are the energy density and
the principal stresses. The unitary vectors $v^{\alpha},~k^{\alpha}$
and $s_{\alpha}$ satisfy the following relations
\begin{equation*}
v^{\alpha}v_{\alpha}=-1,\quad
s^{\alpha}s_{\alpha}=k^{\alpha}k_{\alpha}=1,\quad
s^{\alpha}k_{\alpha}=v^{\alpha}k_{\alpha}=v^{\alpha}s_{\alpha}=0.
\end{equation*}
We define these quantities in comoving coordinates as
\begin{equation}\label{3}
v_{\alpha}=-A\delta^{0}_{\alpha}, \quad
k_{\alpha}=C{\delta}^{2}_{\alpha},\quad
s_{\alpha}={\delta}^{3}_{\alpha}.
\end{equation}

The kinematical variables associated with the nonrotating fluid
distributions are the expansion scalar $\Theta,$ the four
acceleration $a_{\alpha}$ and the shear tensor
$\sigma_{\alpha\beta}$ defined as follows, respectively
\begin{eqnarray}\label{4}
\Theta=v^{\alpha}_{;\alpha}, \quad
a_\alpha=v_{\alpha;\beta}v^\beta,\quad
\sigma_{\alpha\beta}=v_{(\alpha;b)}+a_{(\alpha}v_{\beta)}-
\frac{1}{3}\Theta(g_{\alpha\beta}+v_\alpha{v_\beta}).
\end{eqnarray}
The corresponding expansion scalar and non-vanishing components of
four acceleration and shear tensor are as follows
\begin{eqnarray}\nonumber
\Theta=\frac{1}{A}\left(\frac{\dot{B}}{B}
+\frac{\dot{C}}{C}\right),\quad a_{1}=\frac{A'}{A},\quad
a^2=a^\alpha a_{\alpha}=\left(\frac{A'}{AB}\right)^2,\\\label{5}
\sigma_{11}=\frac{B^{2}}{3A}\left(\frac{2\dot{B}}{B}-\frac{\dot{C}}{C}\right),~
\sigma_{22}=\frac{-C^{2}}{3A}\left(\frac{\dot{B}}{B}-\frac{2\dot{C}}{C}\right),~
\sigma_{33}=\frac{-1}{3A}\left(\frac{\dot{B}}{B}+\frac{\dot{C}}{C}\right).
\end{eqnarray}
Here dot and prime mean differentiation with respect to $t$ and $r$,
respectively.

\subsection{Einstein Field Equations}

The Einstein field equations for Eq.(\ref{1}) yield
\begin{eqnarray}\label{6}
\kappa{\mu}A^{2}&=&\left(\frac{A}{B}\right)^2
\left(\frac{B'C'}{BC}-\frac{C''}{C}\right)
+\frac{\dot{B}\dot{C}}{BC},\\\label{7}
0&=&\frac{\dot{C'}}{C}-\frac{\dot{C}A'}{CA}
-\frac{\dot{B}C'}{BC},\\\label{8}
\kappa{p_r}B^{2}&=&\left(\frac{B}{A}\right)^2
\left(\frac{\dot{A}\dot{C}}{AC}-\frac{\ddot{C}}{C}\right)
+\frac{A'C'}{AC},\\\label{9}
\kappa{p_{\theta}}C^2&=&\left(\frac{C^2}{AB}\right)
\left(\frac{A''}{B}-\frac{\ddot{B}}{A}+\frac{\dot{A}\dot{B}}{A^2}
-\frac{A'B'}{B^2}\right),\\\nonumber
\kappa{p_z}&=&\frac{A''}{AB^2}-\frac{\ddot{B}}{A^2B}
+\frac{\dot{A}\dot{B}}{A^3B}
-\frac{A'B'}{AB^3}+\frac{\dot{A}\dot{C}}{A^3C}
-\frac{\ddot{C}}{A^2C}\\\label{10}
&-&\frac{B'C'}{B^3C}+\frac{C''}{B^2C}
+\frac{A'C'}{AB^2C}-\frac{\dot{B}\dot{C}}{A^2BC}.
\end{eqnarray}
The mass function proposed by Thorne (1965) in the form of
gravitational C-energy per unit specific length of the cylinder is
defined as
\begin{equation}\label{11}
E=\frac{1}{8}(1-l^{-2}{\nabla}^{\beta}\tilde{r}{\nabla}_{\beta}\tilde{r}),
\end{equation}
satisfying the following relations
\begin{equation}\label{12}
\rho^2={\eta_{(\theta)a}}{\eta^a_{\theta}},\quad
l^2={\eta_{(z)a}}{\eta^a_{z}},\quad \tilde{r}=\rho{l},
\end{equation}
where $l,~\rho,~\tilde{r},~\eta_{\theta}$ and $\eta_z$ mean
specific length, circumference radius, areal radius and the
Killing vectors, respectively for the cylindrical geometry. Thus
the specific energy (Poisson 2004) of the collapsing cylinder in
the interior region can be written as
\begin{equation}\label{13}
m(t,r)=El=\frac{l}{8}\left[1+\left(\frac{\dot{C}}{A}\right)^2
-\left(\frac{C'}{B}\right)^2\right].
\end{equation}
The conservation of energy-momentum tensor
($T^{-\alpha\beta}_{~~~~;\beta}=0$) yield the following dynamical
equations
\begin{eqnarray}\label{14}
\dot{\mu}+(\mu+p_r)\frac{\dot{B}}{B}+(\mu+p_\theta)\frac{\dot{C}}{C}=0,
\\\label{14a}
p'_r+(\mu+p_r)\frac{A'}{A}+(p_r-p_\theta)\frac{C'}{C}=0.
\end{eqnarray}
Equation (\ref{14}) with the expansion scalar becomes
\begin{equation}\label{15}
\dot{\mu}+A{\mu}\Theta+\frac{\dot{B}}{B}{p_{r}}+\frac{\dot{C}}{C}p_{\theta}=0.
\end{equation}

\subsection{The Exterior Spacetime and Junction Conditions}

Junction conditions are required to ensure the correct behaviour of
an exterior spacetime source. The outcome of junction conditions
provides the basic information about the pressure on the boundary
surface and total energy entrapped inside the boundary surface. In
case of cylindrical symmetry, there are two possible exteriors
namely the Levi-Civita (Levi-Civita 1917) metric (static case) and
Einstein-Rosen (Einstein and Rosen 1937) metric (non-static case).
Here, we take a timelike $3D$ hypersurface $\Sigma$ which divides
the Riemannian spacetime into two regions interior $M^{-}$ and
exterior $M^{+}$ each containing $\Sigma$ as a part of the boundary.
The junction conditions join these two regions into one across the
surface of discontinuity. We consider two cylindrical regions: the
interior region defined by Eq.(\ref{1}) and the exterior region in
the retarded time coordinate $\nu$ is given by static cylindrical
black hole (Chao-Guang 1995)
\begin{equation}\label{16}
ds^2_+=\left(\frac{2M}{R}\right)d\nu^2
-2d{\nu}dR+R^2(d\theta^2+{\gamma}^2dz^2),
\end{equation}
where $\gamma$ is a constant and has the dimension of
$\frac{1}{r}$ and $M$ is the mass associated with the exterior
geometry.

For the smooth matching of adiabatic cylindrical solution to the
static cylindrical solution, we consider the Darmois conditions
(1927). The continuity of the first and second fundamental forms
yields the following results on the $\Sigma^{(e)}$ (detail is given
in Sharif and Azam 2012a)
\begin{eqnarray}\label{17}
\frac{dt}{d\tau}\overset{\Sigma^{(e)}}=A(t, r)^{-1},\quad
C(t,r)\overset{\Sigma^{(e)}}=R(\nu)=\frac{1}{\gamma},
\\\label{18} \left(\frac{d\nu}{d\tau}\right)^{-2}\overset{\Sigma^{(e)}}=\left(-\frac{2M}{R}
+2\frac{dR}{d\nu}\right),\quad
m-M\overset{\Sigma^{(e)}}=\frac{l}{8}, \quad
p_r\overset{\Sigma^{(e)}}=0.
\end{eqnarray}
The above equations show conditions for the smooth matching of
interior and exterior regions at the boundary $\Sigma^{(e)}$. The
formation of internal vacuum cavity due to the expansion-free
condition within the fluid distribution yields the following
conditions
\begin{equation}\label{17a}
m(t,r)\overset{\Sigma^{(i)}}{=}0,\quad
p_{r}\overset{\Sigma^{(i)}}{=}0,
\end{equation}
where Minkowski spacetime within cavity is matched to the matter
distribution at the internal hypersurface $\Sigma^{(i)}$.

\section{The Perturbation Scheme}

It is well established that any physical model is subject to
$\mu>0$, $\mu>p$ and stability of perturbed modes. In fact,
perturbations are small deviations from static background spacetime
caused by external forces. Here, we use the perturbation scheme
(Sharif and Azam 2012a, 2012b) to perturb the field equations,
dynamical equations, mass function and expansion scalar upto first
order in $\varepsilon$. Initially, we assume that the given fluid
has radial dependence (hydrostatic equilibrium form). Afterwards,
all these functions depend upon $t$. Thus, the metric and material
functions has the following form
\begin{eqnarray}\label{19}
A(t,r)&=&A_0(r)+\varepsilon T(t)a(r),\\\label{20}
B(t,r)&=&B_0(r)+\varepsilon T(t)b(r),\\\label{21}
C(t,r)&=&C_0(r)+\varepsilon T(t)\bar{c}(r),\\\label{22}
\mu(t,r)&=&\mu_0(r)+\varepsilon \bar{\mu}(t,r),\\\label{24}
p_{r}(t,r)&=&p_{r0}(r)+\varepsilon \bar{p_r}(t,r), \\\label{25}
p_{\theta}(t,r)&=&p_{\theta0}(r)+\varepsilon
\bar{p_\theta}(t,r),\\\label{26} m(t,r)&=&m_0(r)+\varepsilon
\bar{m}(t,r),\\\label{26a} \Theta(t,r)&=&\varepsilon
{\bar{\Theta}}(t,r),
\end{eqnarray}
where $0<\varepsilon\ll1$. We choose $C_0(r)=r$ as the radial
coordinate. Using Eqs.(\ref{19})-(\ref{25}), we have static
configuration of the field equations as follows
\begin{eqnarray}\label{27}
\kappa\mu_0&=&\frac{1}{B^2_0}\left(\frac{1}{r}\frac{B'_0}{B_0}\right),\\\label{28}
\kappa{p_{r0}}&=&\frac{1}{B^2_0}\left(\frac{1}{r}\frac{A'_0}{A_0}\right),\\\label{29}
\kappa{p_{\theta0}}&=&\frac{1}{AB_0}\left(\frac{A''_0}{B_0}-\frac{A'_0B'_0}{B^2_0}\right).
\end{eqnarray}
The corresponding perturbed field equations becomes
\begin{eqnarray}\label{30}
\kappa{\bar\mu}&=&\frac{T}{B_0^2}
\left[\frac{B'_0}{B_0}\left(\frac{\bar{c}}{r}\right)'
+\frac{1}{r}\left(\frac{b}{B_0}\right)'-\frac{\bar{c}''}{r}\right]
-2\frac{\kappa{b}}{B_0}\mu_0T,\\\label{30a}
0&=&2\frac{\dot{T}}{A_0B_0}\left[\left(\frac{\bar{c}}{r}\right)'-\frac{b}{rB_0}
-\left(\frac{A'_0}{A_0}-\frac{1}{r}\right)\frac{\bar{c}}{r}\right],\\\label{31}
\kappa\bar{p_r}&=&-\frac{\ddot{T}}{A_0^2}
\left(\frac{\bar{c}}{r}\right)+\frac{T}{B_0^2}
\left[\frac{A_0'}{A_0}\left(\frac{\bar{c}}{r}
\right)'+\frac{1}{r}\left(\frac{a}{A_0}\right)'\right]
-2\frac{{\kappa}b}{B_0}p_{r0}T,\\\nonumber
\kappa\bar{p_\theta}&=&-\frac{\ddot{T}}{A_0^2}
\left(\frac{b}{B_0}\right)+\frac{T}{A_0B_0}
\left[\frac{a''}{B_0}-\frac{A_0B'_0}{B^2_0}\left(\frac{a}{A_0}
\right)'-\frac{aA''_0}{A_0B_0}\right.\\\label{32}
&-&\left.\frac{{A'_0}}{B_0}\left(\frac{b}{B_0}\right)'\right]
-2\frac{{\kappa}b}{B_0}p_{\theta0}T.
\end{eqnarray}

The static and perturbed configurations of the dynamical equations
are
\begin{eqnarray}\label{33}
&&p'_{r0}+(\mu_0+p_{r0})\frac{A_0'}{A_0}+(p_{r0}-p_{\theta0})
\frac{1}{r}=0,\\\label{34}
&&\dot{\bar\mu}+(\mu_0+p_{r0})\frac{b}{B_0}\dot{T}+(\mu_0+p_{\theta0})
\frac{\bar{c}}{r}\dot{T}=0, \\\nonumber
&&{\bar{p_r}}'+(\mu_0+p_{r0})\left(\frac{a}{A_0}\right)'+(\bar{\mu}+\bar{p_r})
\left(\frac{A'_0}{A_0}\right)+(p_{r0}-p_{\theta0})\left(\frac{\bar{c}}{r}\right)'
\\\label{35}&+&\frac{(\bar{p_r}-\bar{p_{\theta}})}{r}=0.
\end{eqnarray}
Integration of Eq.(\ref{34}) yields
\begin{eqnarray}\label{36}
\bar{\mu}=-(\mu_0+p_{r0})\frac{b}{B_0}T-(\mu_0+p_{\theta0})\frac{\bar{c}}{r}T.
\end{eqnarray}
The unperturbed and perturbed configuration of the mass function
give
\begin{eqnarray}\label{37}
m_0=\frac{l}{8}\left[1-\frac{1}{B^2_0}\right], \quad
\bar{m}=-\frac{lT}{4B_0^2}\left[\bar{c}'-\frac{b}{B_0}\right].
\end{eqnarray}
The junction condition (\ref{18}) with Eq.(\ref{24}) provides the
following relations
\begin{equation}\label{37a}
p_{r0}\overset{\Sigma^{(e)}}=0,\quad
\bar{p_{r}}\overset{\Sigma^{(e)}}=0.
\end{equation}
Using these results in Eq.(\ref{31}), it follows that
\begin{equation}\label{38}
\ddot{T}-\alpha{T}\overset{\Sigma^{(e)}}=0,
\end{equation}
where
\begin{eqnarray*}
\alpha(r)&=&\left(\frac{A_{0}}{B_{0}}\right)^2
\left[\frac{A'_0}{A_0}\left(\frac{\bar{c}}{r}\right)'
+\frac{1}{r}\left(\frac{a}{A_0}\right)'
\right]\left(\frac{r}{\bar{c}}\right).
\end{eqnarray*}
The general solution of the above equation becomes
\begin{equation}\label{38a}
T(t)=c_1\exp(\sqrt{\alpha_{\Sigma^{(e)}}}t)
+c_2\exp(-\sqrt{\alpha_{\Sigma^{(e)}}}t),
\end{equation}
where $c_1$ and $c_2$ are arbitrary constants. Here, we are
interested to find a solution which is real and shows a static
system that starts collapsing at $t=-\infty$ when $T(-\infty)=0$. We
assume all the radial functions on the hypersurface to be positive
such that $\alpha_{\Sigma^{(e)}}>0$. The corresponding solution of
Eq.(\ref{38}) has been obtained by setting $c_1=-1$ and $c_2=0$
\begin{equation}\label{39}
T(t)=-\exp(\sqrt{\alpha_{\Sigma^{(e)}}}t).
\end{equation}

\section{Expansion-free Condition and Dynamical Instability}

In this section, we obtain an expansion-free dynamical equation and
investigate the role of anisotropy on the dynamical instability of
collapsing cylinder. For the sake of instability conditions at
Newtonian and pN, we need to distinguish Newtonian, pN and post-post
Newtonian (ppN) terms appearing in the dynamical equation. This has
been done by converting the dynamical equation into c.g.s units.
Inserting the value of $B^2_0$ from Eq.(\ref{37}) in (\ref{27}) and
(\ref{28}), it follows that
\begin{equation}\label{41}
\frac{B_0'}{B_0}=\frac{\kappa{l}\mu_0{r}}{l-8m_0}, \quad
\frac{A_0'}{A_0}=\frac{\kappa{l}{p_{r0}r}}{l-8m_0}.
\end{equation}
Using the value of $\frac{A'_0}{A_0}$ from the above equation in
Eq.(\ref{33}), the dynamical equation in relativistic units becomes
\begin{eqnarray}\label{42}
p_{r0}'=-\left[\frac{\kappa{l}{p_{r0}r}}{l-8m_0}\right](\mu_0+p_{r0})
+\frac{(p_{r0}-p_{\theta0})}{r},
\end{eqnarray}
and in c.g.s. units, it implies that
\begin{eqnarray}\nonumber
p_{r0}'=-G\left[\frac{\kappa{l}{p_{r0}r}}{l-8Gc^{-2}m_0}\right]
(\mu_0+c^{-2}p_{r0})+\frac{(p_{r0} -p_{\theta0})}{r}.\\\label{42a}
\end{eqnarray}
Expanding the above equation up to order $c^{-4}$, we obtain terms
with order $c^0,~c^{-2}$ and $c^{-4}$ corresponding to Newtonian, pN
and ppN order terms given by
\begin{eqnarray}\nonumber
p_{r0}'&=&-{\kappa}Grp_{r0}\mu_0+\frac{(p_{r0}
-p_{\theta0})}{r}-\frac{\kappa{Grp_{r0}}}{c^2}
\left(p_{r0}+\frac{8Gm_0\mu_0}{l}\right)\\\label{42a}
&-&\frac{\kappa{Gp_{r0}}r}{c^4} \left(\frac{8Gm_0p_{r0}}{l}+
\frac{64G^2m^2_0\mu_0}{l}\right).
\end{eqnarray}
The perturbed form of the expansion scalar with Eq.(\ref{26a}) is
\begin{eqnarray}\label{43}
\bar\Theta&=&\frac{\dot{T}}{A_0}\left(\frac{b}{B_0}+\frac{\bar{c}}{r}\right).
\end{eqnarray}
Applying the expansion-free condition $(\bar{\Theta}=0)$, it follows
that
\begin{equation}\label{44}
\frac{b}{B_0}=-\frac{\bar{c}}{r}.
\end{equation}

We would like to mention here that the expansion-free fluids evolve
without being compressed (Herrera et al. 2008) and halt the
formation of apparent horizons resulting a naked singularity (Joshi
et al. 2002). Under the expansion-free condition, Eq.(\ref{30a})
becomes
\begin{equation}\label{44a}
2\frac{\dot{T}}{r^3B_0}\left(\frac{r^2\bar{c}}{A_0}\right)'=0
\quad\Longrightarrow \quad\bar{c}=k_1\frac{A_0}{r^2},
\end{equation}
where $k_1$ is a constant of integration. We relate $\bar{p}_r$
and $\bar{\mu}$ through adiabatic index $\Gamma$ (Herrera et al.
1989)
\begin{equation}\label{45}
\bar{p}_r=\Gamma\frac{p_{r0}}{\mu_0+p_{\theta0}}\bar{\mu},
\end{equation}
where $\Gamma$ measures the stiffness of the fluid and is assumed to
be constant throughout the instability analysis. Using
Eq.(\ref{44}), the perturbed energy density $\bar{\mu}$ takes the
form
\begin{equation}\label{46}
\bar{\mu}=(p_{r0}-p_{\theta0})T\frac{\bar{c}}{r}.
\end{equation}
This shows the relevance of static background anisotropy with the
perturbed energy density, which can be seen through Eq.(\ref{15})
with the expansion-free condition. Inserting this value of
$\bar{\mu}$ in Eq.(\ref{45}), it follows
\begin{equation}\label{47}
\bar{p_r}=2\Gamma\frac{p_{r0}}{\mu_0+p_{r0}}(p_{r0}
-p_{\theta0})T\frac{\bar{c}}{r}.
\end{equation}

We note that $\bar{p}_r$ and $\bar{\mu}\frac{A'_0}{A_0}$ are of ppN
order terms, hence discard these terms in the below stability
analysis. From Eq.(\ref{33}), we obtain
\begin{eqnarray}\label{48}
\frac{A'_0}{A_0}=-\frac{1}{\mu_0+p_{r0}}
\left[p'_{r0}+\frac{(p_{r0}-p_{\theta0})}{r}\right].
\end{eqnarray}
Using the expansion-free condition and value of $\bar{p}_\theta$
from Eq.(\ref{35}) along with Eq.(\ref{39}) in (\ref{32}), the
dynamical expansion-free equation leads to
\begin{eqnarray}\nonumber
&&\kappa(\mu_0+p_{r0})r\left(\frac{a}{A_0}\right)'
+\kappa(p_{r0}-p_{\theta0})r\left(\frac{\bar{c}}{r}\right)'
-\frac{\alpha_{\Sigma^{(e)}}}{A^2_0}\frac{\bar{c}}{r}
-\frac{2\kappa\bar{c}}{r}p_{\theta0}\\\label{49}&-&\frac{1}{A_0B_0}
\left[\frac{a''}{B_0}-\frac{A_0}{B_0}\frac{B'_0}{B_0}
\left(\frac{a}{A_0}\right)'+\frac{A'_0}{B_0}
\left(\frac{\bar{c}}{r}\right)'
-\frac{a}{A_0}\frac{A''_0}{B_0}\right] =0.
\end{eqnarray}
The expressions for $\frac{A''_0}{A_0B^2_0}$ and $(\frac{a}{A_0})'$
from Eqs.(\ref{29}) and (\ref{31}) can be written as
\begin{equation}\label{50}
\frac{A''_0}{A_0B^2_0}=\frac{A'_0B'_0}{A_0B^3_0}+\kappa{p_{\theta0}},
\end{equation}
and
\begin{eqnarray}\nonumber
\left(\frac{a}{A_0}\right)'&=&r\left[\alpha_{\Sigma^{(e)}}
\left(\frac{\bar{c}}{r}\right)\left(\frac{B_0}{A_0}\right)^2
-\frac{A'_0}{A_0}\left(\frac{\bar{c}}{r}\right)'
-2\kappa{B^2_0}p_{r0}\frac{\bar{c}}{r}\right].\\\label{51}
\end{eqnarray}
Inserting Eqs.(\ref{44a}), (\ref{50}) and (\ref{51}) in (\ref{49}),
it yields
\begin{eqnarray}\nonumber
&&\kappa(\mu_0+p_{r0})\left[k_1\frac{\alpha_{\Sigma^{(e)}}}{r}
\left(\frac{B^2_0}{A_0}\right)+3k_1\frac{A_0}{r^2}
\left(\frac{A'_0}{A_0}\right)-2k_1A_0B^2_0\frac{\kappa{p_{r0}}}{r}\right]
\\\nonumber&+&k_1\kappa\frac{(p_{r0}-p_{\theta0})A_0}{r^2}
\left(\frac{A'_0}{A_0}-\frac{3}{r}\right)
-\frac{\alpha_{\Sigma^{(e)}}}{A_0}\frac{k_1}{r^3}
-2k_1\kappa{p_{\theta0}}\frac{A_0}{r^3}-\frac{a''}{A_0B^2_0}
\\\nonumber&+&\frac{1}{B^2_0}\left(\frac{B'_0}{B_0}\right)
\left[k_1\frac{\alpha_{\Sigma^{(e)}}}{r^3}
\left(\frac{B^2_0}{A_0}\right)+3k_1\frac{A_0}{r^4}
\left(\frac{A'_0}{A_0}\right)-2k_1A_0B^2_0\frac{\kappa{p_{r0}}}{r^3}\right]
\\\label{52}&+&\frac{a}{A_0}\left\{\frac{1}{B^2}
\left(\frac{A'_0}{A_0}\right)
\left(\frac{B'_0}{B_0}\right)+\kappa{p_{\theta0}}\right\}=0,
\end{eqnarray}
where we have neglected the ppN terms $\bar{\mu}\frac{A'_0}{A_0}$
and $\bar{p_r}$. Notice that the terms in Eq.(\ref{52}) generally
depend upon the radial function. This shows the relevance of areal
radius in the onset of stability of collapsing fluid.

In order to have instability conditions, we define the following
choice of radial functions $a=a_0+a_1r$, where $a_0,~a_1$ are
arbitrary positive constants. With these choices of radial functions
and metric functions $A_0=1-\frac{Gm_0}{c^2r},~
B_0=1+\frac{Gm_0}{c^2r}$, the dynamical equation (\ref{52}) at pN
approximation becomes
\begin{eqnarray}\nonumber
&&\kappa{\mu_0}\left(1+\frac{3m_0}{r}\right)
\frac{k_1\alpha_{\Sigma^{(e)}}}{r}-\frac{3k_1\kappa}{r^2}
\left(1-\frac{m_0}{r}\right)\left(p'_{r0}+\frac{p_{r0}-p_{\theta0}}{r}\right)
\\\nonumber&-&2\frac{k_1{\kappa}^2p_{r0}\mu_0}{r}
\left(1+\frac{m_0}{r}\right)
-\frac{k_1\kappa(p_{r0}-p_{\theta0})}{r^2}\left(1-\frac{m_0}{r}\right)
\left\{\frac{p'_{r0}}{\mu_0}+\frac{p_{r0}-p_{\theta0}}
{r\mu_0}\right.\\\nonumber&+&\left. \frac{3}{r}\right\}
-\frac{k_1\alpha_{\Sigma^{(e)}}}{r^3}
\left(1+\frac{m_0}{r}+\frac{m^2_0}{r^2}\right)
-2\frac{k_1\kappa{p_{\theta0}}}{r^3}\left(1-\frac{m_0}{r}\right)
+k_1\alpha_{\Sigma^{(e)}}\frac{\kappa{\mu_0}}{r}
\end{eqnarray}
\begin{eqnarray}\nonumber
&\times&\left(1+\frac{m_0}{r}+\frac{8m_0}{l}\right)
-\frac{3k_1\kappa}{r^2}\left(1-\frac{3m_0}{r}+\frac{8m_0}{l}\right)
\left(p'_{r0}+\frac{p_{r0}-p_{\theta0}}{r}\right)
\\\nonumber&-&2\frac{k_1{\kappa}^2\mu_0p_{r0}}{r}
\left(1-\frac{m_0}{r}+\frac{8m_0}{l}\right)-\kappa{r}(a_0+a_1r)
\left(1-\frac{m_0}{r}+\frac{8m_0}{l}\right)
\\\nonumber&\times&\left(p'_{r0}+\frac{p_{r0}-p_{\theta0}}{r}\right)
-\frac{k_1}{r^3}\left(1-\frac{3m_0}{r}\right)
\left(\frac{p'_{r0}}{\mu_0}+\frac{p_{r0}-p_{\theta0}}
{r\mu_0}\right)
\\\label{53}&\times&\left\{\left(\frac{p'_{r0}}{\mu_0}+\frac{p_{r0}-p_{\theta0}}
{r\mu_0}\right)+\frac{3}{r}\right\}
+\kappa{p_{\theta0}}(a_0+a_1r)\left(1+\frac{m_0}{r}\right)=0,
\end{eqnarray}
where we assume $G=1=c$ in the stability analysis. We assume
$\mu_0\gg p_{r0},~\mu_0\gg p_{\perp0}$ and neglect the ppN order
terms to obtain instability conditions at N approximation as
\begin{eqnarray}\nonumber
&-&\kappa\left(6k_1+(a_0+a_1r)r^3\right)rp'_{r0}-k_1\kappa(9p_{r0}-7p_{\theta0})
-\kappa{r^3}(a_0+a_1r)\\\label{54}&\times&(p_{r0}-2p_{\theta0})-k_1\alpha_{\Sigma^{(e)}}
+2k_1\kappa\alpha_{\Sigma^{(e)}}\mu_0r^2-k_1\alpha_{\Sigma^{(e)}}\frac{m_0}{r}=0.
\end{eqnarray}
Using the fact that the radial pressure is decreasing during the
expansion-free collapse, i.e., $p'_{r0}<0$ and Eq.(\ref{13}), the
above equation implies
\begin{eqnarray}\nonumber
&&\kappa\left(6k_1+(a_0+a_1r)r^3\right)r|p'_{r0}|
=k_1\alpha_{\Sigma^{(e)}}+
k_1\kappa(9p_{r0}-7p_{\theta0})\\\label{55}&+&\kappa{r^3}(a_0+a_1r)(p_{r0}-2p_{\theta0})
+2k_1\alpha_{\Sigma^{(e)}}\left(\frac{l\kappa}{8r}
\int^{r}_{r_{{\Sigma}^{(i)}}}\mu_0rdr-\kappa{\mu_0}r^2\right).
\end{eqnarray}

We are interested to find the instability of the expansion-free
fluid at N approximation. For this purpose, we require all the terms
in the above equation positive. Since the quantities $a_0,~a_1$ are
positive constants, thus the instability is subject to the
positivity of each term on the right hand side of (\ref{55}). This
has been assured through $p_{r0}>2p_{\theta0}$ and
$9p_{r0}>11p_{\theta0}$. For the positivity of the last term, we
consider power law solution for the energy density, i.e.,
$\mu_0=\zeta{r^n}$, where $\zeta>0$ is a constant and $n$ has range
in the interval $(-\infty,\infty)$. The last term of the above
equation for $n\neq-2$ implies
\begin{equation}\label{57}
\frac{l\kappa}{8r}\int^{r}_{r_{{\Sigma}^{(i)}}}\mu_0rdr
-\kappa{\mu_0}r^2=\frac{l\kappa{\zeta}}{8r(n+2)}
\left(r^{n+2}-\frac{8(n+2)}{l}r^{n+3}-r^{n+2}_{{\Sigma}^{(i)}}\right),
\end{equation}
which reduces to
$r(1-\frac{8r(n+2)}{l})^\frac{1}{n+2}>r_{{\Sigma}^{(i)}}$. For
$n=-2$, it gives
\begin{equation}\label{58}
\frac{l\kappa}{8r}\int^{r}_{r_{{\Sigma}^{(i)}}}\mu_0rdr
-\kappa{\mu_0}r^2=\frac{l\kappa{\zeta}}{8r} \left[\log
\left(\frac{r}{r_{{\Sigma}^{(i)}}}\right)-\frac{8r}{l}\right],
\end{equation}
yielding $re^\frac{-8r}{l}>r_{{\Sigma}^{(i)}}$. These two
inequalities describe the ranges of instability of the
expansion-free fluid, which generally depends upon the given value
of $n$ and specific length of the cylinder.

\section{Summary}

In this paper, we have explored the issue of dynamical instability
of expansion-free fluid with cylindrical symmetry. We have assumed
collapsing cylinder with inhomogeneous energy density and locally
anisotropic pressure, which is compatible with expansion-free
models. The matching conditions at the internal hypersurface
(separating fluid distribution to cavity) and external hypersurface
(dividing the vaccum solution to fluid) have been formulated. We
have used perturbation scheme to distinguish the Newtonian, pN and
ppN terms.

The relevance of expansion-free models in the study of
astrophysical objects stems from the fact that these models may be
helpful in the evolution of cosmic voids. Voids are underdense
regions that fill $40-50\%$ volume of the universe. The
application of such a study would be seen in the cosmological
phenomena where evolution of cavity in the given fluid
distribution has been the subject of interest (Randall et al.
2011). Generally, the adiabatic index determines the dynamical
instability of self-gravitating objects. For instance, the
isotropic cylinder and sphere are unstable for $\Gamma<1$ and
$\Gamma<\frac{4}{3}$, respectively (Sharif and Azam 2012a;
Chandrasekhar 1964). We have shown that Eq.(\ref{55}) is
independent of the adiabatic index, which supports the fact that
expansion-free collapse continues without contraction of the
fluid. Hence, the adiabatic index is irrelevant in this study. In
our case, we have found that the stability of the fluid is
affected by the local anisotropy of pressure and energy density.

Recently, we have explored the dynamical instability of
expansion-free fluid with spherically symmetry and found that the
dynamical instability is subject to $p_{r0}>\frac{2}{5}p_{\perp0}$
and has range in the interval $(-2,2)$ (Sharif and Azam 2012b). In
this work, we have found that the instabilities of Eq.(\ref{55})
at Newtonian regime correspond to the constraints
$p_{r0}>2p_{\theta0},~9p_{r0}>7p_{\theta0}$,
$r(1-\frac{8r(n+2)}{l})^\frac{1}{n+2}>r_{{\Sigma}^{(i)}}$ and
$re^\frac{-8r}{l}>r_{{\Sigma}^{(i)}}$. This shows that the
instability of the fluid has no particular range and depends upon
the value of $n$ and the specific length of the cylinder.

Moreover, these inequalities describe the instability ranges of
expansion-free fluid as well as the associated cavity with the
fluid. The violation of any of the inequality would lead to diminish
the instability. From these results, it is evident that physical
properties of fluid, i.e., pressure anisotropy and energy density
play a vital role in the onset of dynamical instability, which
supports the fact that these quantities are important in studying
the dynamics of self-gravitating objects. Finally, we remark that
the stability analysis of the expansion-free fluid would be the same
at pN approximation, where the relativistic correction terms
$\frac{m_0}{r}$ have been taken into account.

\vspace{0.5cm}

{\bf Acknowledgments}

\vspace{0.5cm}

We would like to thank the Higher Education Commission, Islamabad,
Pakistan, for its financial support through the {\it Indigenous
Ph.D. 5000 Fellowship Program Batch-VII}. One of us (MA) would like
to thank University of Education, Lahore for the study leave.

\end{document}